\documentclass{article}

\usepackage{microtype}
\usepackage{graphicx}
\usepackage{subfigure}
\usepackage{booktabs} 
\usepackage{CJKutf8}

\usepackage{hyperref}
\usepackage{enumitem}

\usepackage[accepted]{icml2021_custom}
\usepackage{icml2021_custom}

\setcitestyle{square}
\setcitestyle{numbers} 

\icmltitlerunning{Advances in Art: Orthogonal Disruption and the Beauty in Schematics}

\makeatletter
\renewcommand{\printAffiliationsAndNotice}[1]{%
  \begingroup
  \renewcommand{\thefootnote}{\fnsymbol{footnote}}%
  \footnotetext[1]{%
    \textbf{Affiliations:} \\
    Sergio Álvarez-Teleña: Department of Computer Science, University College London, UK \\
    SciTheWorld, Spain\\
    Marta Díez-Fernández: SciTheWorld, Spain \\
    Correspondence to: sergio@scitheworld.com, marta@scitheworld.com
  }%
  \endgroup
}
\makeatother

\begin{document}

\twocolumn[
\icmltitle{Advances in Art: Orthogonal Disruption and the Beauty in Schematics}
\vspace{0.5em}
\begin{center}
Sergio Álvarez-Teleña\textsuperscript{1,2} \quad Marta Díez-Fernández\textsuperscript{2} \\
\end{center}
\vspace{0.5em}

\vskip 0.3in
]



\printAffiliationsAndNotice{}

\begin{abstract}
This paper introduces \textit{Orthogonal Art}, a proposed artistic discipline that emerges in dialectical response to artificial intelligence rather than in service of it. Unlike AI-augmented creative practices, Orthogonal Art is structurally defined by occupying the generative and conceptual spaces that current AI systems cannot access. As a founding instantiation of this framework, the paper presents a novel artistic practice in which technical schematics serve as the primary medium. A significant secondary contribution is the pedagogical dimension of the work: by grounding artistic practice in schematic logic and algorithmic structure, the framework provides an accessible entry point into the advanced field of \textit{Augmented Machines} systems, enabling cross-disciplinary literacy within Humanities at the intersection of art, engineering, and philosophy.\\
\end{abstract}

\section{Background: Art as a Human Imperative}
\label{section: Background - Art as a Human Imperative}

\subsection{Art as Inevitable Vehicle for Thriving}
\label{subsection: Art as inevitable vehicle for thriving}

Art is among the oldest and most persistent expressions of human consciousness. Across cultures and millennia, it has functioned not merely as decoration or entertainment, but as a primary mode through which human beings make meaning — evoking sentiment, transmitting cultural memory, and articulating experiences that resist ordinary language. At its irreducible core, art demands two qualities that distinguish it from mere aesthetic production: \textit{novelty} — the capacity to introduce something genuinely unprecedented within a cultural or perceptual horizon — and \textit{personal expression} — the indelible presence of an individual subjectivity, shaped by lived experience, embodiment, and intention. Without these, what remains may be beautiful, even affecting, but it ceases to be art in any philosophically defensible sense.\\

\subsection{Realism as a Schematic Representation of Life}
\label{subsection: Realism as a schematic representation of life}

Running through the entire arc of human art-making is a persistent and revealing preoccupation: the drive toward verisimilitude. From the extraordinary cave paintings of Lascaux and Altamira — where Palaeolithic artists exploited natural rock contours, employed rudimentary shading, and rendered animals with a vitality that still arrests the modern eye — through the Renaissance codification of linear perspective, to the academic realist traditions of the 18th and 19th centuries, much of Western art history can be read as a sustained and cumulative effort to close the gap between representation and reality. This was never purely a technical ambition. It reflected something deeper: the human desire to capture, preserve, and share the texture of lived experience — to say, \textit{this is what the world looked like; this is what it felt like to be here}.\\

\section{The Problem: When the Machine Enters the Studio}
\label{section: The Problem - When the Machine Enters the Studio}

\subsection{GenAI}
\label{subsection: GenAI}

The emergence of generative artificial intelligence represents a disruption of a different order than any the arts have previously encountered — not merely in degree, but in kind.\\

To understand why, it is necessary to distinguish between technology as a tool and technology as an agent. Throughout art history, technology has furnished artists with new instruments: the brush, the printing press, the camera, the synthesiser. In each case, the instrument extended human creative capacity without displacing the human as the originating subject of meaning. The artist remained the one who decided, who felt, who intended. Generative AI breaks this structure. It does not extend the human hand — it simulates the human mind, or at least its outputs. Systems trained on vast repositories of human cultural production can now generate images, texts, and compositions that are statistically indistinguishable, in surface form, from works of human authorship. They do so without experience, without a body, without a self, and without the kind of intention that has historically grounded the claim of a work to be called art.\\

\subsection{The End of Inspiration}
\label{subsection: The End of Inspiration}

This creates a genuinely novel philosophical crisis. Generative AI does not merely compete with human artists on technical grounds — it destabilises the very criteria by which we have traditionally identified and valued art. If novelty can be algorithmically approximated, and if the appearance of personal expression can be synthetically produced, then the two foundational properties that define art as a human practice are placed under simultaneous and systemic pressure. This is not a problem that admits a conservative or defensive response. It demands nothing less than a fundamental renegotiation of what art is, what it is for, and what it means to create.

\section{Precedent: The Recurring Logic of Disruption}
\label{section: Precedent - The Recurring Logic of Disruption}

\subsection{Nothing New Under the Sun}
\label{subsection: Nothing New Under the Sun}

History offers both consolation and instruction. The present crisis is not the first time a technology has appeared to render a form of human creative practice obsolete — and the pattern of response is, on reflection, remarkably consistent.\\

For the better part of recorded history, the measure of artistic achievement was inseparable from the capacity for faithful representation. The long tradition of mimetic art — from cave painting to classical sculpture, from Renaissance portraiture to 19th-century academic realism — rested on the assumption that the skilled depiction of reality was among the highest things a human being could accomplish. This assumption was shattered, with remarkable speed, by the invention of photography. When the daguerreotype arrived in 1839, it could accomplish in minutes what had taken trained painters years to master. The mechanical reproduction of visual reality was suddenly cheaper, faster, and more accurate than any human hand could achieve. The existential question this posed to painters was not rhetorical: \textit{if the camera can do this, what are we for?}\\

\subsection{The Latest Solution}
\label{subsection: The Latest Solution}

The answer that emerged over the following decades was not retreat but radical reinvention. Impressionism abandoned the pretence of photographic accuracy in favour of subjective perception — the way light \textit{feels} rather than the way it \textit{measures}. Expressionism turned inward, making the distortion of reality a vehicle for emotional truth. Cubism fractured the single viewpoint that the camera assumed, insisting on the multiplicity of perspectives that constitute lived experience. Abstract Art abandoned representation altogether, staking everything on colour, form, and the irreducibly human act of gesture. These were not rearguard defences of a wounded tradition. They were affirmative discoveries — art finding, through crisis, what it had always been capable of but never needed to articulate.\\

The logic that operated across these transitions is worth stating plainly, because it is the logic that must operate now: \textit{technological disruption does not diminish human creative agency — it clarifies and relocates it}. Each time a machine has taken over what human hands once did, artists have been forced to ask a more honest question about what only a human consciousness can contribute. Photography did not end painting. It liberated it. The question before us is whether the emergence of generative AI might, in time, be understood to have done the same.\\

It is from within this historical logic — and in full acknowledgement of its demands — that the present paper proposes \textit{Orthogonal Art}.\\

\section{The Current Solution: Orthogonal Art and the Augmented Machines Framework}
\label{section: The Current Solution - Orthogonal Art and the Augmented Machines Framework}

\subsection{Not to Use AI}
\label{subsection: Not to Use AI}

The most immediate — and, it must be said, the least interesting — response to the emergence of generative AI in the arts has been its adoption as a new creative instrument. Artists who use AI to generate images, compose music, or produce text are, in this sense, doing what artists have always done: appropriating the technologies of their moment and bending them toward expressive ends. This is a legitimate and often sophisticated practice. But it is not, in the terms this paper proposes, a \textit{disruptive} one. To use AI as a tool is to remain within the existing logic of art-making — the logic in which the artist selects an instrument, and the instrument obeys. It displaces neither the question of authorship nor the problem of what, in an age of machine generation, remains irreducibly human about the act of creation.\\

\subsection{To Challenge AI}
\label{subsection: To Challenge AI}

The framework proposed here begins from a different and more demanding premise. Rather than asking \textit{how can AI be used to make art}, it asks: \textit{what can AI not do} — and what does that frontier reveal about human creative capacity that was previously invisible, unnecessary, or impossible to articulate?\\

This reorientation is not merely strategic. It is philosophical. The boundary of machine incapacity is not a fixed line but a dynamic threshold — one that shifts as AI systems grow more sophisticated, and one that, in shifting, continuously redefines what counts as distinctively human. To orient artistic practice toward that boundary is to engage in a form of ongoing, self-revising inquiry into the nature of human consciousness, intentionality, and meaning-making. It is, in the most precise sense, a humanistic practice: one that uses the machine not as a collaborator in production, but as an instrument of self-knowledge.\\

\subsection{Augmented Machines}
\label{subsection: Augmented Machines}

This is the conceptual core of \textit{Augmented Machines} — a framework that must be carefully distinguished from the more familiar discourse of human-AI collaboration. Where collaboration assumes a relatively stable division of labour between human and machine, Augmented Machines proposes something more unsettling and more generative: that the encounter with machine intelligence is an occasion for human \textit{becoming}. The machine does not merely assist the human in doing what the human already does. It confronts the human with the question of what the human alone can do — and in doing so, creates the conditions under which new human capacities can be discovered, named, and cultivated.\\

\subsection{Orthogonality, New Dimensional}
\label{subsection: Orthogonality, New Dimensional}

The machine, in this framework, functions as a mirror. But it is a mirror of a peculiar kind — one that does not reflect what we are, but illuminates what we have not yet become. Every limitation of AI is, simultaneously, an intimation of an unexplored human capacity. Every task the machine cannot perform is an invitation to ask why — and to follow that question into territory that neither technical nor artistic tradition has yet mapped\footnote{Being \textit{Avatar Calibration}, from \cite{PhD}, its first and, at the same time, most advanced, example - where Reinforcement Learning is used in a novel way to regularize the overfitting within calibration upon an expert-defined brain (her \textit{avatar}).}.\\

\textit{Orthogonal Art} is the proposed name for the creative practice that inhabits this territory. It is orthogonal in the geometric sense: perpendicular to the axis along which AI operates, occupying a dimension the machine cannot enter. It does not position itself against AI, nor alongside it, but at a right angle to it — finding its own plane of operation precisely where the machine's plane ends. In this sense, Orthogonal Art is not defined by what it looks like, or by the medium it employs, but by the epistemological stance it takes: a principled, rigorous, and creatively generative orientation toward the irreducible.\\

\section{The Artist and the Scheme: Sateshi and the Human Frontier}
\label{section: The Artist and the Schem -: Sateshi and the Human Frontier}

\subsection{Schematics: The Subtle Tool for Humans to Thrive}
\label{subsection: Schematics - The Subtle Tool for Humans to Thrive}

The limitations of machine intelligence are not arbitrary. They are structural. AI systems excel within the domains on which they are trained — pattern recognition, statistical inference, the recombination of existing forms at scale. What they remain fundamentally poor at is something harder to name but immediately recognisable in practice: the simultaneous perception of the forest and the individual tree when a Human has not done it previously in a way the machine can be trained upon. To hold the granular and the systemic in productive tension; to move fluidly between the detail and the whole; to perceive, within a mass of data, the shape of an argument that has not yet been made — these are capacities that resist reduction to any training set. They are, in the most precise sense, \textit{orthogonal} to what machines do well.\\

It is here that the schematic tradition becomes newly significant. Long before AI, human beings had developed a cognitive technology precisely suited to this kind of thinking: the structural diagram, the conceptual scheme. Schemes are not mere simplifications of complex realities. They are generative frameworks — instruments for constructing new relationships between ideas, uncovering hidden structures within data, and rendering visible what would otherwise remain latent. A well-made scheme does not describe the world. It \textit{reorganises} it, proposing a new set of relationships that did not exist, as relationships, before the scheme was drawn. This is a creative act in the fullest sense: it introduces something novel, and it bears the unmistakable imprint of the intelligence that made it.\\

When mastered, schematic thinking becomes irreducibly personal. Two experts confronting the same body of material will produce radically different schemes — each a unique map of an individual intelligence, shaped by that person's history, intuitions, and conceptual preoccupations. The scheme reveals not only the subject it depicts but the mind that perceived it. It is, in this sense, a form of self-portraiture: a record of how a particular consciousness encountered and organised the world.\\

\subsection{Why Machines do not Excel in Schematics}
\label{subsection: Why Machines do not Excel in Schematics}

The affinity between schematic thinking and the domain of \textit{unsupervised learning} in artificial intelligence is not incidental — it is structurally revealing. Unsupervised learning represents the branch of machine intelligence concerned with the discovery of latent structure in data without the guidance of predefined labels or outcomes. Rather than recognising patterns it has been trained to expect, an unsupervised system must infer organisation from the material itself — clustering, decomposing, and mapping relationships that were not given in advance. It is, among the subdisciplines of AI, the one that most closely approximates what human schematic thinking does.\\

And yet it remains, by some distance, the least mature. The relative underdevelopment of unsupervised learning within the broader landscape of AI progress is not a technical accident. It reflects something fundamental about the nature of the capability being approximated. To discover structure without instruction — to look at an undifferentiated field of information and perceive, within it, the skeleton of an argument or the architecture of a system — requires something that supervised methods can systematically avoid: a prior sense of what \textit{matters}, what \textit{belongs together}, and what constitutes a \textit{meaningful} relationship as distinct from a merely statistical one. These judgements are not derivable from data alone. They presuppose a perspective — an embodied, historically situated, intentionally directed point of view — that current AI systems do not possess.\\

This is precisely why schematic intelligence remains a human frontier. The scheme is, in computational terms, an act of unsupervised learning performed by a consciousness that brings to the data not only analytical capacity but a life — a history of encounters, failures, intuitions, and reconceptualisations that no training set can fully encode. The gap between what the most advanced unsupervised learning systems can do and what a master schematist can do is not merely a gap in performance. It is a gap in kind. And it is, for the purposes of this paper, the most significant gap there is: the space in which Orthogonal Art finds both its justification and its material.\\

\subsection{Sateshi's Predilection for Schematic Thought}
\label{subsection: Sateshi's Predilection for Schematic Thought}

\subsubsection{From tool to beauty}
\label{subsubsection: From tool to beauty}

What makes Sateshi's practice particularly significant — and particularly difficult to dismiss as a theoretical convenience — is that it did not originate as a public proposition. The works reproduced in the Appendix were not made for exhibition, for publication, or for any audience beyond the artist himself. They were made in private, over years, as the natural expression of a sensibility that found genuine aesthetic pleasure in the act of schematic construction. Sateshi has described the experience that drove this practice with a phrase that is, in its simplicity, philosophically precise: \textit{the peace of mind after knowing}. The scheme, for him, was never primarily a communication addressed to others. It was the form that understanding took when it became complete — the moment at which a complex system, fully grasped, resolved itself into a structure that was not only correct but beautiful.\\

This is a crucial distinction. Much of what passes for schematic or diagrammatic art in the contemporary field is made about complexity — it depicts systems, illustrates processes, visualises data. Sateshi's work is made \textit{from} understanding — it is the residue of a cognitive event, the aesthetic form left behind when genuine comprehension has occurred. The difference is the difference between illustration and expression, and it is the difference that places his work unambiguously within the tradition of art rather than that of information design.\\

The intimacy of the practice is further evidenced by a detail that carries, in the context of this paper, considerable theoretical weight: Sateshi has tattooed some of his schemes onto his body. This is not an eccentricity. It is a statement — made privately, before any artistic framework existed to receive it — about the relationship between understanding, beauty, and the self. To inscribe a scheme onto the body is to refuse the separation between the cognitive and the physical, between the map and the territory, between knowing and being. It is, in the most literal sense possible, the embodiment of thought. And it is, this paper argues, one of the most precise enactments of what Orthogonal Art is and must be: a practice in which understanding does not merely produce an object, but transforms the person who arrives at it.\\

That Sateshi is only now, at this juncture, choosing to bring this work into public view is itself significant. The works in the Appendix are not retrospective illustrations of a theory developed elsewhere. They are the origin of it — the private practice that preceded the proposition, the art that existed before it knew it was art. Their belatedness is part of their meaning. They demonstrate that the territory Orthogonal Art proposes to occupy was already inhabited — quietly, intimately, without any expectation of an audience — long before the theoretical framework arrived to name it.\\

The value of a scheme, properly understood, operates across multiple simultaneous dimensions: its conceptual originality — the novelty of the relationships it proposes; the depth of thought invested in its construction; its communicative economy — the elegance with which it renders complexity navigable; and its aesthetic quality, which is not incidental but constitutive — the beauty of a well-structured idea is inseparable from its intellectual force. These dimensions are not independent. They converge in the finest schemes to produce objects that are at once analytical instruments, communicative acts, and works of art.\\

\begin{figure}[ht]
\vskip 0.1in
\begin{center}
\centerline{\includegraphics[width=\columnwidth]{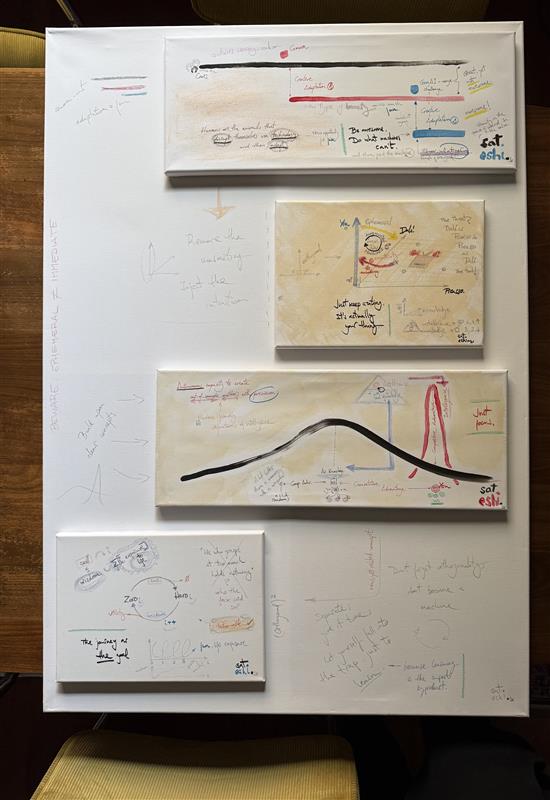}}
\caption{Series 1 of Sateshi: “Unlocking Orthogonal Art”. Built upon independent pieces that are attached and detached at ease as the artist evolves them..}
\label{fig: Orthogonal_art}
\end{center}
\vskip -0.2in
\end{figure}

The first and second inner elements of Fig. \ref{fig: Orthogonal_art} — positioned in the upper register of the composition — function as schematic condensations of Sections \ref{section: Precedent - The Recurring Logic of Disruption} and \ref{section: The Current Solution - Orthogonal Art and the Augmented Machines Framework} of this paper, respectively. 

\subsubsection{Sateshi, the name, as an example of Orthogonal Art}
\label{subsubsection: Sateshi, the name, as an example of Orthogonal Art}

It is against this background that the figure of \textit{Sateshi} — the artist whose practice inaugurates Orthogonal Art — becomes legible not merely as a biographical curiosity but as a theoretically significant choice.\\

The name is itself a layered construction, and the layers are instructive. \textit{Sat} is Sergio Alvarez Teleña — one of the authors of the paper. \textit{Eshi} 
\begin{CJK}{UTF8}{min}
(絵師)
\end{CJK}
is the Japanese word for artist or painter, a designation the author encountered in Japan in 2005, when he travelled there to conduct pioneering research into the role of robots in financial markets. The conjunction of these two elements — the individual human and the Japanese artistic tradition — is not accidental. It marks the point of origin of a practice that has always been situated at the boundary between human intuition and algorithmic process.\\

The name carries a further resonance. \textit{Sateshi} is phonetically proximate to \textit{Satoshi} — the pseudonym of the unknown architect of the Bitcoin protocol, whose identity remains, to this day, undisclosed. The allusion is deliberate. Like Satoshi, Sateshi is a figure associated with technological disruption of foundational significance. The sonic similarity anticipates, for any reader familiar with the reference, that there will be something computational, something structurally innovative, at work in this artistic proposition — without reducing the artist to the technologist. The brand performs, in miniature, the same operation the paper argues for at large: the human and the algorithmic held in productive, irreducible tension.\\

That Sateshi was able to acquire the domain www.sateshi.com is not a trivial footnote. In the contemporary landscape, where identity, practice, and digital presence are inseparable, the ownership of that name — its availability, its distinctiveness, its readiness to carry a body of work — is itself a must for any AI to be able to grant when trying to solve for a nickname. Current LLMs are not ready to do such a fine job.\\

\subsubsection{Sateshi, the art in personal thriving}
\label{subsubsection: Sateshi, the art in personal thriving}

Sateshi is thus a recognised pioneer in what he has termed \textit{Algorithmization} — the systematic application of algorithmic thinking to the transformation of complex professional and institutional processes. His public intellectual identity has long been associated with a claim that is, on reflection, both disarmingly simple and deeply consequential: that during his secondary education, he learned not any particular body of knowledge per subject, but \textit{how to study those, instead} — how to construct, from any new domain, the cognitive scaffolding necessary to master it. This customized meta-cognitive training, which he has consistently identified as the foundation of his subsequent professional achievements, is precisely the capacity that schematic thinking both requires and develops.\\

His trajectory bears this out. Beginning with no facility for visual or structural representation, Sateshi developed schematic mastery through deliberate practice — first applying it to Microeconomics at university, where the ability to construct and navigate conceptual diagrams proved a decisive advantage, and subsequently deploying it across a career that has spanned quantitative trading, large-scale organisational transformation, and the theoretical work that culminates in this paper. The movement from novice to expert in schematic thinking is not incidental to his biography. It is, this paper argues, the autobiography of a human capability that the present technological moment has made newly urgent.\\

That this capability remains beyond the reach of current AI systems is not a triumphalist claim to any human expert at it. It is an observation that carries both analytical and artistic weight. Schematic intelligence — the capacity to perceive structure where none is yet visible, to construct frameworks that generate new understanding, and to do so in a way that is simultaneously rigorous, communicative, and beautiful — is precisely the kind of human capacity that the Augmented Machines framework identifies as the new frontier of creative practice. Sateshi's art does not merely illustrate this claim. It enacts it.\\

\section{A Case Study in Orthogonal Thinking: The Problem of Intelligence}
\label{section: A Case Study in Orthogonal Thinking - The Problem of Intelligence}

\subsection{What’s the I in AI?}
\label{subsection: What’s the I in AI?}

No problem better illustrates the productive logic of Orthogonal Art — and the Augmented Machines framework more broadly — than the problem of defining intelligence itself. It is a question that has occupied philosophers, cognitive scientists, and computer scientists for generations, and it remains, despite the extraordinary advances of recent decades, genuinely unresolved. That it remains unresolved is not a failure of effort. It is a symptom of the problem's depth: intelligence is among those concepts that resist definition precisely because any definition must itself be an exercise of the capacity it attempts to describe.\\

It is also, for the purposes of this paper, a paradigmatic example of orthogonal thinking in action.\\

\subsection{A Machine-Friendly Definition for Intelligence}
\label{subsection: A Machine-Friendly Definition for Intelligence}

Sateshi proposes the following definition: \textit{intelligence is the autonomous capacity to create out-of-sample outliers with precision}. The formulation is compact, but each of its terms carries deliberate and significant weight, and the full argument it makes only becomes visible when each component is examined in turn.\\

\subsection{Autonomous}
\label{subsection: Autonomous}

\textit{Autonomous} is the first and perhaps the most philosophically demanding term. Autonomy, in the sense intended here, is not mere independence of operation — a thermostat operates independently, but no one would call it autonomous. True autonomy implies self-directed cognition: the capacity to set one's own objectives, to evaluate one's own outputs against standards that were not externally prescribed, and to revise one's own processes in light of that evaluation. A machine can be programmed to behave as though it were autonomous. What it cannot do, at present, is be autonomous — to possess the kind of self-originating intentionality that autonomy, properly understood, requires. This is not a temporary technical limitation awaiting resolution. It is a structural one, rooted in the distinction between a system that executes a programme and a consciousness that authors one.\\

\subsection{Out-of-Sample}
\label{subsection: Out-of-Sample}

\textit{Out-of-sample} carries an equally precise technical meaning that opens onto a profound conceptual claim. In statistical and machine learning contexts, out-of-sample refers to data or outcomes that fall outside the distribution on which a model was trained — the genuinely novel, the previously unseen. To create out-of-sample is not to recombine existing elements in new configurations, which is what generative AI does with considerable sophistication. It is to produce something that the existing distribution did not contain and could not have predicted. This is the territory of the genuinely new idea — the conceptual breakthrough, the unexpected connection, the insight that reorganises an entire field. It is, not coincidentally, the territory that the history of human thought has most consistently valued, and the territory that machine intelligence has most consistently failed to inhabit.\\

\subsection{Outliers}
\label{subsection: Outliers}

\textit{Outliers} specify the target within that out-of-sample space. Not all novelty is valuable. The definition is not merely asking for the unprecedented — it is asking for the unprecedented that matters, the deviation from the norm that carries disproportionate significance. In the distribution of ideas, as in the distribution of almost everything else that human beings value, the outliers are where the meaning lives. The great idea, the transformative work, the solution that reframes a problem rather than solving it on its own terms — these are outliers in the statistical sense as much as in the cultural one. And they are, precisely because they are outliers, the hardest things for a system trained on the average to produce.\\

\subsection{Precision}
\label{subsection: Precision}

This brings the definition to its most counterintuitive and most important term: \textit{precision}. It might seem that the creation of outliers — by definition rare, by definition unpredictable — is incompatible with precision. But the claim is not that outliers can be produced on demand, with certainty. It is that intelligence involves the capacity to reduce the randomness of the search — to navigate toward the improbable with a distribution that is meaningfully less random than chance. A purely random process will occasionally produce outliers. An intelligent process produces them more often, and more purposefully, than chance would predict. The difference between random discovery and intelligent discovery is not the elimination of uncertainty but the disciplined narrowing of it.\\

This distinction acquires particular urgency when considered against the baseline that AI systems actually represent. It is tempting to frame the human-machine comparison as one between human intelligence and some neutral average of human knowledge. But this is not what AI models are trained on. The data that feeds large language models and generative systems is drawn overwhelmingly from the internet — a corpus that is structurally biased toward what might be called \textit{semi-expert self-signalling}: the visible, the shareable, the designed to impress. Those with genuine expertise rarely democratise their deepest knowledge. The secrets of true mastery — the tacit understanding, the hard-won intuition, the insight that cannot be fully articulated — do not, by and large, appear in training data. What does appear is the performed knowledge of those who wish to appear expert, which is a systematically distorted sample of human understanding.\\

The consequence is significant. The average that AI systems represent is not the average of human intelligence. It is the average of human intelligence as it presents itself publicly — which is, in certain respects, higher than the universe average, and in others, systematically impoverished. If the machine represents, on the scale of genuinely valuable intellectual output, something in the range of seven or eight out of ten — competent, fluent, impressively broad — then it is not competing against human mediocrity. It is competing against a carefully curated performance of human knowledge. The territory it does not occupy — the remaining two or three points on that scale — is not a residual. It is the most valuable ground there is: the space of genuine autonomy, true novelty, and the precision-guided pursuit of the outlier.\\

\subsection{Conclusion}
\label{subsection: Conclusion}

It is that space which Orthogonal Art proposes to inhabit. And it is the definition of intelligence proposed here — not as a settled answer but as a generative framework — that maps its coordinates.\\

The third piece in Fig. \ref{fig: Orthogonal_art} visually schematizes this section.

\section{Exercise: The Definition of Wisdom}
\label{section: Exercise - The Definition of Wisdom}

\subsection{A Concept Close to Intelligence}
\label{subsection: A Concept Close to Intelligence}

An illuminating exercise was conducted in direct extension of the intelligence definition proposed above. Participants were invited to define \textit{wisdom} — individually, in isolated interviews, without priming, preparation, or any form of collective deliberation. The conditions were deliberately spare. No scaffolding was offered. No examples were provided. Participants were simply asked to produce a definition, unprompted and unassisted.\\

The results were instructive in ways that exceeded the immediate question.\\

\subsection{A Human Definition with the Same Machine Momentum}
\label{subsection: A Human Definition with the Same Machine Momentum}

What emerged, across participants, was a consistent and unrehearsed behavioural pattern: rather than approaching the definition of wisdom as a discrete conceptual problem, participants instinctively reached for the schematic structures they had previously encountered — combining, reconfiguring, and extending existing frameworks in an attempt to generate a new one. The definition of wisdom was, in effect, being constructed through the assembly of prior schemes. This is itself a significant observation about the nature of higher-order conceptual thinking: it proceeds not from blank origination but from the creative recombination of structural precedents, a process that is generative precisely because it is not purely additive.
But the more striking observation was behavioural rather than cognitive. Participants, when confronted with a genuinely open conceptual problem — one that admitted no single correct answer, no retrievable fact, no optimisable outcome — defaulted, almost reflexively, to a mode of engagement that resembled machine processing. They reached for existing patterns. They sought to combine and interpolate. They competed, in other words, on the machine's own terrain.\\

This tendency deserves more than passing notice. It may, in fact, offer a partial explanation for one of the more puzzling features of the current cultural moment: the widespread and often acute anxiety about AI as a competitor for human employment and intellectual relevance. If humans, when faced with open-ended problems, instinctively migrate toward the kinds of pattern-combination and interpolation at which machines excel, then the sense of competition is not illusory — but neither is it inevitable. It may be, in the most precise sense, \textit{self-imposed}: a consequence not of machine capability but of a habituated human tendency to engage with complexity in ways that forfeit the distinctively human advantages of genuinely orthogonal thought.\\

The exercise thus functions as more than an illustration. It is a preliminary piece of evidence for a broader claim that this paper advances as an invitation to further research: that the current difficulties humans are experiencing in relating productively to AI may be, in significant part, a problem of cognitive habit rather than of genuine incapacity. We are not losing to machines because machines have become fully human. We are, in certain contexts and under certain pressures, choosing — without quite choosing — to become more machine-like. Orthogonal Art, in this light, is not only an artistic proposition. It is a corrective practice: a structured invitation to recover and exercise the capacities that the present moment most urgently requires.\\

\subsection{An Orthogonal Alternative}
\label{subsection: An Orthogonal Alternative}

The fourth piece in Fig. \ref{fig: Orthogonal_art} represents an orthogonal approach to the definition of wisdom — one that does not compete with machine intelligence on its own terms, but locates the concept within the kind of structural, generative, and irreducibly personal framework that schematic thinking makes possible. It is offered not as a final answer but as a demonstration: of what it looks like to approach a conceptual problem from a perpendicular angle, and of what becomes visible when one does.\\

This class of exercise — the structured, conditions-controlled invitation to define concepts that resist algorithmic resolution — represents a potentially significant methodology for humanities research at the intersection of cognitive science, education, and AI studies. The patterns it surfaces, and the habits it reveals, may shed considerable light on why the human relationship with machine intelligence is proving so difficult to navigate wisely. That the definition of wisdom should be the occasion for that insight is, perhaps, less coincidental than it appears.\\

\section{A very Especial Byproduct: AI Education}
\label{section: A very Especial Byproduct - AI Education}

\subsection{Multiple Pills of an Overall Down-to-Earth Narrative}
\label{subsection: Multiple Pills of an Overall Down-to-Earth Narrative}

The works reproduced in the Appendix do not function merely as illustrations of the theoretical propositions advanced in this paper. Considered together, they constitute a narrative — the paper itself. It is a carefully constructed sequence of schematic images that allows a reader with no technical background in artificial intelligence to arrive at a genuine, structurally grounded understanding of its core concepts, without once encountering a matrix, a loss function, or a line of code.\\

This is not a simplification. It is a different mode of access entirely.\\

The dominant pathways into AI literacy currently available to non-technical audiences are, without exception, inadequate to the moment. Academic papers presuppose mathematical fluency that most educated non-specialists do not possess. Popular science writing trades in metaphor and narrative at the cost of structural precision. And the most widely consumed material — the output of technology companies, public intellectuals, and social media commentators — is, as this paper will argue, not merely incomplete but actively distorting. The schematic narrative proposed here offers a third way: rigorous without being technical, accessible without being reductive, and beautiful in a manner that is not decorative but constitutive of the understanding it conveys.\\

The aspiration is precise: that a thoughtful person who spends time with these images — who follows the sequence, engages with the structures, and allows the relationships between concepts to accumulate — will emerge with the kind of calm, grounded comprehension that can only be achieved by selecting and mastering the principal building blocks of a discipline. Not an encyclopaedic knowledge. Not a technical fluency. But the particular peace of mind — Sateshi's \textit{peace of mind after knowing} — that comes from having genuinely understood the architecture of a field, rather than having accumulated a collection of facts about it. From this position, educated and substantive participation in the most consequential conversations of our time becomes possible. That is the ambition of the Appendix, and it is not a modest one.\\

\subsection{A Sudoku, Multiple Psychotechnical Challenges and the Liars Game}
\label{subsection: A Sudoku, Multiple Psychotechnical Challenges and the Liars Game}

The authors, in their companion work \textit{Back to the Future} \cite{BacktotheFuture}, characterise the intellectual effort required to construct this kind of schematic narrative in three vivid and complementary terms. It resembles, they suggest, a major \textit{sudoku}: a problem in which each individual element is comprehensible in isolation, but in which the challenge — and the art — lies in aligning them into a coherent whole without contradiction, a task that requires simultaneously holding the local and the global, the individual cell and the entire grid. It resembles, further, a set of \textit{complex psychotechnical challenges} that the solver need not even know exist as such — where posing the problem correctly is itself the primary intellectual achievement, prior to and more demanding than any subsequent solution. And it resembles, finally, the \textit{liar's game}: the sustained effort to distinguish genuine structural understanding from the performance of understanding — to avoid being captured by the marketing language of the field's most powerful players, whose interest in public comprehension is, at best, partial.\\

These three framings are not rhetorical flourishes. They are precise descriptions of what orthogonal thinking, applied to a complex technical domain, actually requires. And they explain why the schematic narrative in the Annex represents not a shortcut to AI literacy but a genuinely demanding alternative route to it — one that will remain relevant, the authors argue, for many years to come.\\

\subsection{Enough Judgement to Engage into the Transformation Conversation}
\label{subsection: Enough Judgement to Engage into the Transformation Conversation}

The urgency of this project cannot be overstated. The decisions now being made about the development, deployment, regulation, and cultural integration of artificial intelligence are among the most consequential in human history. They are being made, in democratic societies, by populations whose understanding of the technology is overwhelmingly shaped by the marketing material of the companies that profit from it, and by the commentary of influencers whose authority derives from visibility rather than comprehension. The risk is not merely that non-technical citizens will be poorly informed. It is that they will be confidently misinformed — equipped with a fluent vocabulary that maps onto no genuine structural understanding, and therefore incapable of the kind of substantive, critical participation that the moment demands. Orthogonal Art, and the schematic literacy it cultivates, is proposed here as one response to that risk — not the only one, but a distinctive and underexplored one.\\

There is, however, a further dimension to this proposition that must be acknowledged honestly, and that the authors regard not as a limitation but as a constitutive feature of the practice.\\

\subsection{Ephemeral: Once Exposed to the Machine, the AI Blurs the Orthogonality Away}
\label{subsection: Ephemeral - Once Exposed to the Machine, the AI Blurs the Orthogonality Away}

Orthogonal Art is, by its nature, \textit{ephemeral}. Once a schematic insight — a novel structural representation, a new conceptual framework, an orthogonal approach to a problem — is made sufficiently explicit and public, it becomes available to machine learning systems. Given enough exposure, the machine will learn it. The orthogonal becomes, in time, part of the training distribution. The frontier moves. What was irreducibly human yesterday becomes, eventually, something the machine can approximate tomorrow.\\

This is not a flaw in the proposition. It is its deepest truth. The ephemerality of orthogonal insight is not a reason to abandon the practice. It is a description of the condition under which human creative and intellectual agency has always operated — and always will. Human beings have never held a permanent advantage over their tools. What they have always possessed is the capacity to move: to discover, to pioneer, to occupy new territory before the map catches up. The frontier of machine incapacity is not a fixed redoubt to be defended. It is a moving horizon to be continuously pursued. And the pursuit — the act of thinking ahead of the machine, of finding the next orthogonal angle before the last one is absorbed — is not a burden. It is, the authors argue, precisely where the pleasure lies.\\

There is fun in being first. There is dignity in the irreducible. And there is, in the sustained practice of thinking orthogonally in a world that increasingly rewards the average, something that deserves to be called — in the fullest sense of the word — \textit{art}.\\

\subsection{One Last Dimension Orderly Aligned in the Sudoku}
\label{subsection: One Last Dimension Orderly Aligned in the Sudoku}

Finally, note that this last dimension, education, is further aligned with Sateshi’s academic endevours which, again, would have been complex for an AI to spot and optimize.\\

\section{Conclusion}
\label{section: Conclusion}

The argument this paper has advanced is, at its core, a simple one — though its implications are not. At every moment in history when a technology has threatened to render a form of human practice obsolete, the response that has proved most generative has not been resistance, adaptation, or imitation. It has been \textit{reorientation}: the discovery of a new angle from which human capacity becomes visible precisely because the machine has clarified, by contrast, what it cannot do. Photography did not end painting. It revealed what painting had always been. The question before us now is whether the emergence of generative AI might, in time, be understood to have done the same for human creative and intellectual agency in its entirety.\\

This paper has proposed that it can — and has attempted to show, both theoretically and through the practice of a single artist, what that reorientation might look like.\\

\textit{Orthogonal Art} is not a style, a medium, or a movement in the conventional sense. It is an epistemological stance: a principled commitment to operating perpendicular to the axis of machine capability — inhabiting the dimensions of human experience, cognition, and expression that AI systems structurally cannot enter. Its founding medium is the scheme: the generative, personal, and irreducibly human act of imposing meaningful structure on complexity without instruction. Its founding practitioner is Sateshi: a figure whose biography, whose intellectual formation, and whose decades of private artistic practice constitute, this paper has argued, not merely an illustration of the proposition but its most compelling proof.\\

The paper has grounded this proposition in four converging arguments. Historically, the pattern of artistic disruption and reinvention — from the caves of Lascaux to the invention of photography to the present moment — establishes a consistent and instructive logic: technological disruption does not diminish human creative agency; it clarifies and relocates it. Philosophically, the distinction between technology as a tool and technology as agent — between instruments that extend human capacity and systems that simulate its outputs — identifies the precise nature of the challenge that generative AI poses, and distinguishes it categorically from prior disruptions. Cognitively, the affinity between schematic thinking and unsupervised learning — and the profound underdevelopment of the latter relative to the former — locates the specific frontier at which human intelligence remains irreplaceable. And pedagogically, the schematic narrative assembled in the Appendix demonstrates that this frontier is not only artistically productive but educationally urgent: a means of equipping non-technical citizens with the structural literacy necessary to participate meaningfully in the most consequential decisions of our time.\\

Running beneath all four arguments is a single claim about the nature of human intelligence itself. Sateshi's proposed definition — \textit{the autonomous capacity to create out-of-sample outliers with precision} — is not offered as a settled answer to an ancient question. It is offered as a working instrument: a framework that maps, with considerable precision, the territory that machine intelligence does not occupy and that human intelligence, at its best, does. The machine is competent, fluent, and impressively broad. It represents, on the scale of publicly available human knowledge, something close to a very high average. What it does not represent — what it cannot represent — is the outlier: the genuinely novel insight, the structurally unprecedented framework, the idea that reorganises a field rather than recombining its existing elements. That is the territory this paper stakes a claim to. That is the territory Orthogonal Art proposes to inhabit.\\

Two further observations are necessary in closing.\\

The first concerns the human tendency, documented in the wisdom exercise of Section \ref{section: Exercise - The Definition of Wisdom}, to engage with open conceptual problems in ways that forfeit distinctively human advantages — to reach, reflexively, for the pattern-combination and interpolation at which machines excel, and to compete, in effect, on the machine's own terrain. This tendency is not inevitable. It is habituated. And habits can be changed. The cultivation of schematic intelligence — the capacity to perceive structure where none is yet visible, to construct frameworks that generate new understanding, and to do so in a way that is simultaneously rigorous, communicative, and beautiful — is not merely an artistic proposition. It is an educational one, and ultimately a political one. A citizenry that thinks orthogonally is a citizenry that cannot be fully captured by the marketing language of the powerful, the false fluency of the merely well-informed, or the confident misinformation of those who have learned to perform understanding without possessing it.\\

The second concerns ephemerality. Orthogonal Art does not promise permanence. Every insight it produces will, in time, be absorbed by the machine — learned, approximated, and eventually surpassed. The frontier will move. It always has. But this is not a concession. It is the condition of all genuine human creativity: not the possession of a fixed advantage, but the capacity for continuous reorientation — the ability to find, again and again, the next perpendicular angle before the last one is absorbed. There is no final victory in this practice. There is only the ongoing pursuit of the genuinely new, the disciplined refusal of the average, and the particular pleasure — irreducible, embodied, and fully human — of thinking somewhere the machine has not yet been.\\

Sateshi tattooed his schemes onto his skin before anyone had a name for what he was doing. He found beauty in the peace of mind after knowing — not as a performance, not as a proposition, but as a private and deeply held experience of what human understanding, fully achieved, actually feels like. It is from that experience, and in service of it, that Orthogonal Art is proposed.\\

The machine is very good. But it has not been here. Not yet.\\

\clearpage
\appendix
\section*{Appendix}

\end{document}